\documentclass{Interspeech}

\newif\ifready

\readytrue

\interspeechcameraready

\usepackage{appendix}
\usepackage{comment}
\usepackage{amsmath,amssymb,amsfonts}
\usepackage{algorithmic}
\usepackage{graphicx}
\usepackage{textcomp}
\usepackage{xcolor}

\usepackage{hyperref}
\usepackage{booktabs}

\newcommand{\A}{\overline{A}}
\newcommand{\B}{\overline{B}}

\title{aTENNuate: Optimized Real-time Speech Enhancement with Deep SSMs on Raw Audio}

\author[affiliation={1}]{Yan Ru}{Pei}
\author[affiliation={1}]{Ritik}{Shrivastava}
\author[affiliation={1}]{FNU}{Sidharth}

\affiliation{}{BrainChip}{USA}
\email{yanrpei@gmail.com, rshrivastava@brainchip.com, sidharth@brainchip.com}

\usepackage{comment}

\begin{document}

\maketitle

\begin{abstract}
We present aTENNuate, a simple deep state-space autoencoder configured for efficient online raw speech enhancement in an end-to-end fashion. The network's performance is primarily evaluated on raw speech denoising, with additional assessments on tasks such as super-resolution and de-quantization. We benchmark aTENNuate on the VoiceBank + DEMAND and the Microsoft DNS1 synthetic test sets. The network outperforms previous real-time denoising models in terms of PESQ score, parameter count, MACs, and latency. Even as a raw waveform processing model, the model maintains high fidelity to the clean signal with minimal audible artifacts. In addition, the model remains performant even when the noisy input is compressed down to 4000Hz and 4 bits, suggesting general speech enhancement capabilities in low-resource environments.
\end{abstract}

\section{Introduction}
\label{sec:intro}

Speech enhancement (SE) is crucial for improving both human-to-human communication, such as in hearing aids, and human-to-machine communication, as seen in automatic speech recognition (ASR) systems. A challenging task of SE is removing background noises from speech signals, where the complexity of speech and noise patterns poses challenges \cite{germain2018speech}. Traditional speech enhancement methods such as Wiener filtering \cite{chen_new_2006}, spectral subtraction \cite{vaseghi_spectral_1996}, and principal component analysis \cite{srinivasarao_speech_2020} have shown satisfactory performance in stationary noise environments, but their effectiveness is often limited in non-stationary noise scenarios, resulting in artifacts and substantial degradation in the intelligibility of the enhanced speech \cite{yu_speech_2020}.

Deep learning audio denoising methods are trained on large datasets of clean and noisy audio pairs, and attempt to capture the nonlinear relationship between the noisy and clean signal features without prior knowledge of the noise statistics, as required by traditional denoising methods \cite{azarang_review_2020}. Speech features can be extracted from real or complex spectrograms of the noisy signal in the time-frequency domain, or directly from raw waveforms \cite{zhao_convolutional_2019}. 

Many state-of-the-art deep learning denoising models leverage the feature extraction capability of convolutional networks. The UNet convolutional encoder-decoder is a common network architecture in denoising models 
\cite{giri_attention_2019, defossez_real_2020, kashyap_speech_2021} 
exemplified by Deep Complex UNet \cite{choi_phase-aware_2019} and the generative adversarial SEGAN \cite{pascual_segan_2017}. However, CNNs lack the capabilities to model long-range temporal dependencies present in speech signals, without incurring significant memory resources \cite{wavenet}. Another class of models that are more capable in this regard are RNNs, examples of which are DCCRN \cite{hu_dccrn_2020} and FRCRN \cite{frcrn}. However, these models typically show limited robustness to different types of noise and generalization across diverse audio sources \cite{zheng_sixty_2023}. Moreover, their non-linear recurrent nature prevents efficient usage of parallel hardware (e.g. GPUs) for training, which limits their potential for scaling.


Alternative approaches such as PercepNet \cite{percepnet} and RNNoise \cite{rnnnoise} have attempted to reduce network size and complexity by combining traditional speech enhancement methods with deep learning, resulting in smaller models with fewer parameters capable of running in real-time on general-purpose hardware. Most notably, DeepFilterNet \cite{dfn} has leveraged speech-specific properties such as short-time speech correlations to achieve comparable results. On the contrary, methods that process raw waveform signals aim to maximize the expressive capabilities of deep networks without resorting to hand-engineered spectral conversions, as demonstrated by Facebook's DEMUCS model's real-time performance \cite{demucs}. However, the majority of these models cannot perform real-time inference on general-purpose hardware, exemplified by Nvidia's CleanUNet model \cite{kong_speech_2022}.

Here, we introduce the aTENNuate network, belonging to the class of Temporal Neural Networks \cite{tenn_eye, pleiades, centaurus}, or TENNs. It is a deep state-space model (SSM) \cite{s4, sashimi} optimized for real-time denoising of raw speech waveforms on the edge. By virtue of being an SSM, the model is capable of capturing long-range temporal relationships present in speech signals, with stable linear recurrent units. Learning long-range correlations can be useful for capturing global speech patterns or noise profiles, and perhaps implicitly capture semantic contexts to aid speech enhancement performance \cite{selm}. During the training of the aTENNuate network, we can use the infinite impulse response (IIR) kernels of the SSM layers as long convolutional kernels over the input features, which can be parallelized using techniques such as FFT convolution \cite{s4d} or associative scan \cite{s5}. During inference, the temporal convolution layers can be converted into equivalent recurrent layers for efficient real-time processing on mobile devices, minimizing latency and reducing the need for excessive buffering of data. The evaluation code is \href{https://pypi.org/project/attenuate/}{available on PyPI}, installable with \texttt{pip install attenuate}.

\section{Deep State-space Modeling}
\label{ssm}

\begin{figure}[htbp]
    \centering
    \includegraphics[width=\linewidth]{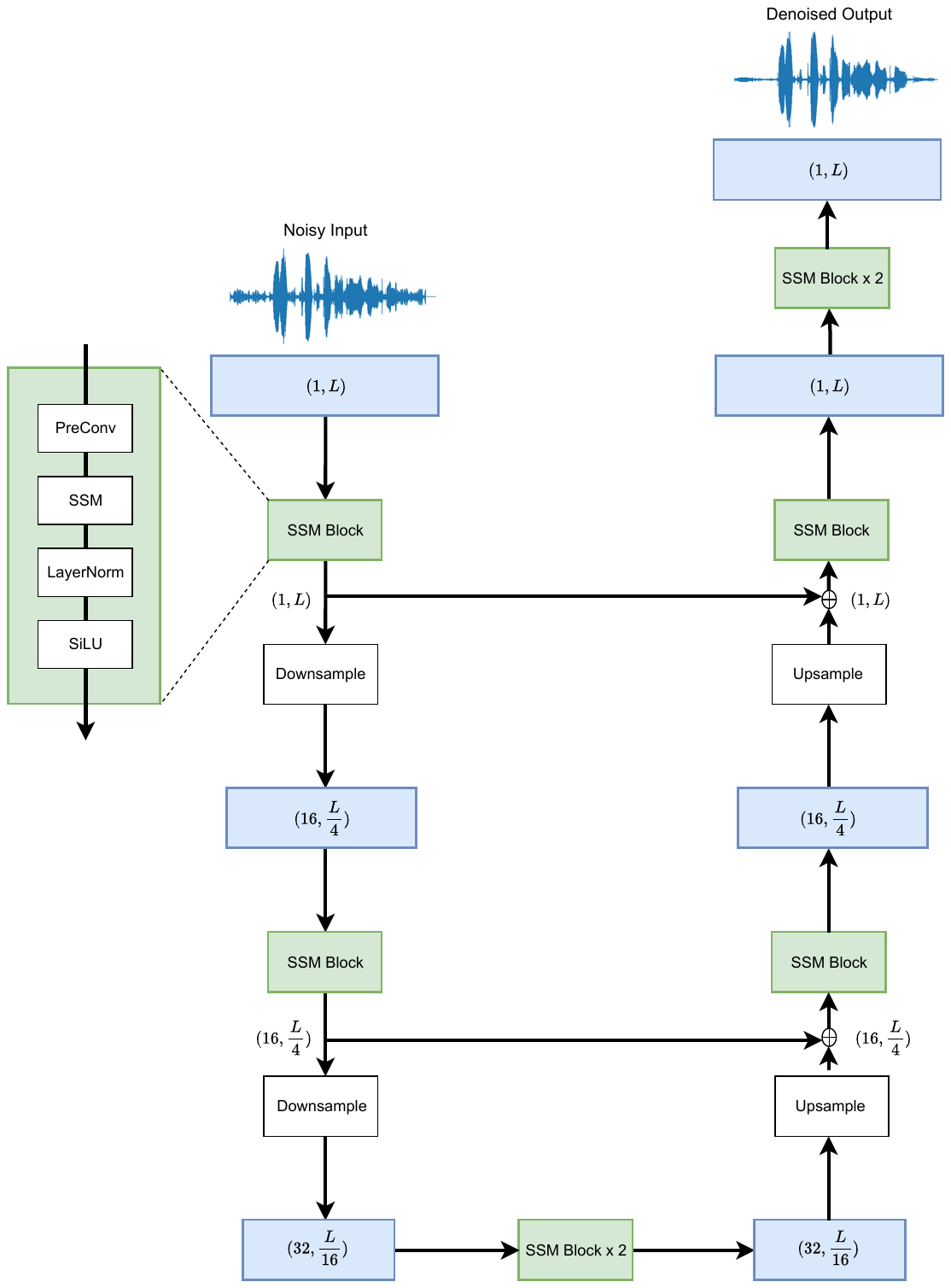}
    \caption{\label{fig:network} A schematic drawing of the network architecture, with only 2 encoder and decoder blocks shown for simplicity. The actual model has 6 encoder and decoder blocks. Note that there is no (spectral) processing on the input and output waveforms.}
\end{figure}

In this section, we briefly describe what state-space models (SSMs) are, and recall how they can be configured for neural network processing, involving discretization and diagonalization. SSMs are general representations of linear time-invariant (LTI) systems, and they can be uniquely specified by four matrices: $A \in \mathbb{R}^{h \times h}$, $B \in \mathbb{R}^{h \times n}$, $C \in \mathbb{R}^{m \times h}$, and $D \in \mathbb{R}^{m \times n}$. The first-order ODE describing the LTI system is given as
\begin{equation}
\dot{x} = Ax + Bu, 
\qquad
y = Cx + Du,
\end{equation}
where $u \in \mathbb{R}^n$ is the input signal, $x \in \mathbb{R}^h$ is the internal state, and $y \in \mathbb{R}^m$ is the output. Here, we are letting $n>1, \, m>1$, which yields a multiple-input, multiple-output (MIMO) SSM. For the remainder of this paper, we will ignore the $Du$ term as it effectively serves as a skip connection \cite{s4d}, which is already included explicitly in our network (see Fig.~\ref{fig:network}).

The SSM in its original form describes a continuous-time system, but in the field of digital signal processing, there are standard recipes for discretizing such a system into a discrete-time SSM. One such method that we use in this work is the zero-order hold (ZOH), which gives us the discrete-time state-space matrices $\A$ and $\B$ as follows:
\begin{equation}
\A = \exp(\Delta A), 
\qquad
\B = (\Delta A)^{-1} \cdot (\exp(\Delta A) - 1) \cdot \Delta B.
\end{equation}
The discrete SSM is then given by
\begin{equation}
x[t+1] = \A\,x[t] + \B\,u[t], 
\qquad
y[t] = C \,x[t]
\end{equation}
In other words, this is essentially a linear RNN layer, which allows for efficient online inference and generation (in our case real-time speech enhancement), but at the same time efficient parallelization during training.

It is straightforward to check that the discrete-time impulse response is given as
\begin{equation}
k[\tau] = C \, \A^{\tau} \, \B = C K(\tau) \B,
\end{equation}
where $\tau$ denotes the kernel timestep and $K$ represents the ``basis kernels''. During training, $k$ can then be considered the ``full" long 1D convolutional kernel with shape, in the sense that the output $y$ can be computed via the long convolution $y_j = \sum_{i} u_i \ast k_{ij}$. By the convolution theorem, we can perform this operation in the frequency domain, which becomes a point-wise product $\hat{y}_{jf} = \sum_i \hat{u}_i \hat{k}_{ijf}$. The hat symbol denotes the Fourier transform of the signal (with the index $f$ denoting the Fourier modes), which can be efficiently computed via Fast Fourier Transforms (FFTs).

It is a generic property that a diagonal form exists for the SSM, meaning that we can almost always assume $\A$ to be diagonal \cite{s4d}, at the expense of potentially requiring $\B$ and $C$ to be complex matrices. Since the original system is a real system, the diagonal $\A$ matrix can only contain real elements and/or complex elements in conjugate pairs. 
\ifready
    (See Appendix \ref{appendix_label}.)
\fi
In this work, we sacrifice a slight loss in expressivity by continuing to restrict $\B$ and $C$ to be real matrices and letting $\A$ be a diagonal matrix with all complex elements (but not restricting them to come in conjugate pairs). Since we still want to work with real features\footnote{This is {\it a prior} not required, as technically we can configure our network as complex-valued to handle complex features. However, we do not explore this configuration in this work.}, we then only take the real part of the impulse response kernel as such:
\begin{equation}
k[\tau] = \Re(C \A^{\tau} \B),
\end{equation}
which equivalently in the state-space equation can be achieved by simply letting $y[t] = C \, \Re(x[t])$. This means that during online inference, we need to maintain the internal states $x$ as complex values, but only need to propagate their real parts to the next layer.

\subsection{Optimal Contractions}

Similar to previous works in deep SSMs, we allow the parameters $\{A, B, C, \Delta\}$ to be directly learnable, which indirectly trains the kernel $k$. Unlike previous works in deep SSMs, we do not try to keep the sizes $n$, $h$, $m$ homogeneous, to allow for more flexibility in feature extraction at each layer, mirroring the flexibility in selecting channel and kernel sizes in convolutional neural networks\footnote{The size of the internal state $h$, can be interpreted as the degree of parametrization of a basis temporal kernel, or some implicit (dilated) ``kernel size" in the frequency domain. We explore this in a future work.}. The flexibility of the tensor shapes requires us to carefully choose the optimal order of operations during training to minimize the computational load.
\ifready
    This is also briefly discussed in Appendix~\ref{app:network}.
\fi

In short, in einsum notation, our SSM operations can be written as $\hat{y}_{jf} = \hat{u}_{if}B_{in} \hat{K}_{nf} C_{jn}$, a series of tensor contractions (matrix multiplications). Based on the tensor shapes involved, we can select an order of operation that is memory- and compute-optimal \cite{centaurus, pleiades}. In addition, this allows us to freely configure the network as an hourglass network (with the SSM layers in the network having different channel and temporal dimensions), without enforcing the SISO form \cite{sashimi,spiking_s4,s4nd_se}. This makes our network uniquely capable of handling raw audio waveforms, while retaining strong featural interactions and light computational costs. See supplementary material or our sister work Centaurus \cite{centaurus} for details on optimal tensor contractions.

\section{Network Architecture}
\label{sec:network}

We use an hourglass network with long-range skip connections, similar in form to the Sashimi network \cite{sashimi} for audio generation. However, unlike previous works using SSMs for audio processing \cite{sashimi,spiking_s4,s4nd_se,se_mamba,mamba_speech}, our network directly takes in raw audio waveforms in the -1 to +1 range and outputs raw waveforms as well, with no one-hot encoding or spectral processing (e.g. STFT or iSTFT). Furthermore, we retain causality as much as possible for the sake of real-time inference, meaning that we eschew any form of bidirectional state-space layers. See Fig.~\ref{fig:network} for a schematic drawing.

As with typical autoencoder networks, the audio features are down-sampled in the encoder and then up-sampled in the decoder. For the re-sampling operation, we use a simple operation that squeezes/expands the temporal dimension then projects the channel dimension \cite{sashimi}. More formally, for a reshaping ratio of $r$, a sequence of features can be down-sampled and up-sampled as:
\begin{equation}
\begin{split}
\text{Down:}& \,\, (C_{in},L) \xrightarrow{\text{reshape}} (C_{in}r,L/r) \xrightarrow{\text{project}} (C_{out},L/r) \\
\text{Up:}& \,\, (C_{in},L) \xrightarrow{\text{reshape}} (C_{in}/r,Lr) \xrightarrow{\text{project}} (C_{out},Lr)
\end{split}
\end{equation}

The baseline network uses LayerNorm layers and SiLU activations. In addition, we include a ``PreConv" layer which is a depthwise 1D convolution layer with a kernel size of 3, to enable better processing of local temporal features. The PreConv operation is omitted in the 2 SSM blocks in the neck and in any block with only one channel\footnote{The neck operates on fully-downsampled features, and introducing PreConv layers will incur too much latency in real-time processing. We also omit PreConv layer in single-channel blocks, because it is too lossy.}. 
\ifready
    See Appendix~\ref{app:network} for details on the computation of the theoretical latency for a given set of resampling factors and PreConv configurations.
\fi
Using causal convolutions for PreConvs can eliminate additional latencies to the network \cite{tenn_eye}, but we do not explore this implementation here. To better support mobile devices, we also test variants of the network with BatchNorm layers\footnote{Since BatchNorm is a static form of normalization during inference, the normalization statistics and the affine parameters can be ``folded" into the weights and biases of the previous layer, meaning that the layer does not need to be materialized during inference.}, ReLU activations, and omitting PreConv layers in the decoder (and omitting PreConv layers altogether). The results are reported in the next section. The block-wise network architecture is given in Table~\ref{tab:network}, along with the PreConv latency, if present.

\begin{table}[htbp]
    \centering
    \caption{\label{tab:network} Resampling factor and output channel of each block of the network, which consists of an encoder performing down-samplings, an intermediate bottleneck, a decoder performing up-samplings, and finally an output processor.}
    \begin{tabular}{lccc}
        \toprule
        \textbf{Layers} & \textbf{Resample Factor} & \textbf{Channels} & \textbf{Latency} \\
        \midrule
        \textbf{Encoder} &  &  & \\
        \quad Block 1 & 4 & 16 & \\
        \quad Block 2 & 4 & 32 & 0.25ms \\
        \quad Block 3 & 2 & 64 & 1ms \\
        \quad Block 4 & 2 & 96 & 2ms \\
        \quad Block 5 & 2 & 128 & 4ms \\
        \quad Block 6 & 2 & 256 & 8ms \\
        \midrule
        \textbf{Neck} &  &  & \\
        \quad Block 1 & 1 & 256 & \\
        \quad Block 2 & 1 & 256 & \\
        \midrule
        \textbf{Decoder} &  &  & \\
        \quad Block 1 & 2 & 128 & 8ms \\
        \quad Block 2 & 2 & 96 & 4ms \\
        \quad Block 3 & 2 & 64 & 2ms \\
        \quad Block 4 & 2 & 32 & 1ms \\
        \quad Block 5 & 4 & 16 & 0.25ms \\
        \quad Block 6 & 4 & 1 & \\
        \midrule
        \textbf{Output} &  &  & \\
        \quad Block 1 & 1 & 1 & \\
        \quad Block 2 & 1 & 1 & \\
        \bottomrule
    \end{tabular}
\end{table}

\section{Experiments}

\begin{table*}[htbp]
    \centering
    \caption{\label{table:main} Comparing different variants of the aTENNuate network against other real-time audio denoising networks, in terms of performance, memory/computational requirements, and latency.}
    \begin{tabular}{lccccc}
        \toprule
        \textbf{Model} & \textbf{PESQ (VB-DMD)} & \textbf{PESQ (DNS1 no-reverb)} & \textbf{Parameters} & \textbf{MACs / sec} & \textbf{Latency} \\
        \midrule
        FRCRN \cite{frcrn} & 3.21 & {\bf 3.23} & 6.9M & 38.1G & 30ms \\
        DeepFilterNet3 \cite{dfn} & 3.16 & 2.58 & 2.13M & 0.344G & 40ms \\
        DEMUCS \cite{demucs} & 2.56 & 2.65 & 33.53M\textsuperscript{a} & 7.72G\textsuperscript{a} & 40ms \\
        PercepNet \cite{percepnet} & 2.73\textsuperscript{b} & - & 8.00M & 0.80G & 40ms \\
        \midrule
        {\bf aTENNuate} (base) & {\bf 3.27} & 2.98 & 0.84M & 0.33G & 46.5ms \\
        \hspace{2mm} PreConvs only in encoder & 3.21 & 2.84 & 0.84M & 0.33G & 31.25ms \\
        \hspace{2mm} no PreConvs & 3.06 & 2.59 & 0.84M & 0.33G & {\bf 16ms} \\
        \hspace{4mm} BatchNorm + ReLU & 2.84 & 2.43 & {\bf 0.84M} & {\bf 0.33G} & {\bf 16ms} \\
        \bottomrule
    \end{tabular}
    \\
    \noindent\footnotesize{
    \textsuperscript{a} These numbers are estimated by passing a one-second segment of data to the model. \\
    \textsuperscript{b} This metric is taken directly from the paper as an official implementation of the model does not exist. \\
    }
\end{table*}

For experiments, we train on the VCTK and LibroVox training sets downloaded from the Microsoft DNS Challenge, randomly mixed with noise samples from Audioset, Freesound, and DEMAND. We evaluate our denoising performance on the Voicebank + DEMAND (VB-DMD) testset, and the Microsoft DNS1 synthetic test set (with no reverberation) \cite{dns1}. To guarantee no data leakage between the training and testing sets, we removed the clean and noise samples in the training set that were used to generate the synthetic testing samples. Both the input and output signals of the network are set at 16000 Hz. The loss function is a mix of SmoothL1Loss \cite{fast_rcnn} and spectral loss at the ERB scale \cite{dfn}. 

We train the model for 500 epochs, AdamW optimizer with a learning rate of 0.005 and a weight decay of 0.02, augmented with a cosine decay scheduler with a linear warmup of 0.01 of the total training steps. Each epoch contains the full VCTK training set, a random subset of the LibroVox training set (of ratio 0.1). To synthesize random noisy samples on the fly as inputs to the network, the clean samples are mixed randomly with the noise samples, with SNR values uniformly sampled from -5 dB to 15 dB.
\ifready
    More details of the training pipeline are provided in Appendix~\ref{app:exp_details}. 
\fi

The evaluation metric is the average wideband PESQ score between the clean signals and the denoised outputs. The PESQ scores for different variants of the aTENNuate model against other real-time audio-denoising networks are reported in Table.~\ref{table:main}. We also report other network inference metrics including parameters, MACs, and latencies. For the latency metric, we focus on the theoretical latency of the network, not accounting for any processing latencies. In simpler terms, it is the maximum time range that the network needs to ``wait" or look-forward to produce a denoised data point corresponding to the current input. As seen in Table~\ref{table:main}, the PreConv layers (being depthwise) does not make much difference in terms of parameters and MACs, but does add considerably to latency. Unless otherwise denoted, we use the source code and pre-trained weights of each model and run it through a standardized PESQ evaluation pipeline. In addition, we report results for other common speech-enhancement metrics in Table~\ref{table:metrics}.

\begin{table}[htbp]
    \centering
    \caption{\label{table:metrics} Various speech enhancement metrics for the base aTENNuate network.}
    \begin{tabular}{lccccc}
        \toprule
        \textbf{Testset} & \textbf{PESQ} & \textbf{CSIG} & \textbf{CBAK} & \textbf{COVL} & \textbf{SISDR} \\
        \midrule
        \textbf{VBDMD} & 3.27 & 4.57 & 2.85 & 3.96 & 15.04 \\
        \midrule
        \textbf{DNS1} & 2.98 & 4.28 & 3.55 & 3.57 & 15.40 \\
        \bottomrule
    \end{tabular}
\end{table}

\begin{figure}[htbp]
    \centering
    \includegraphics[width=\linewidth]{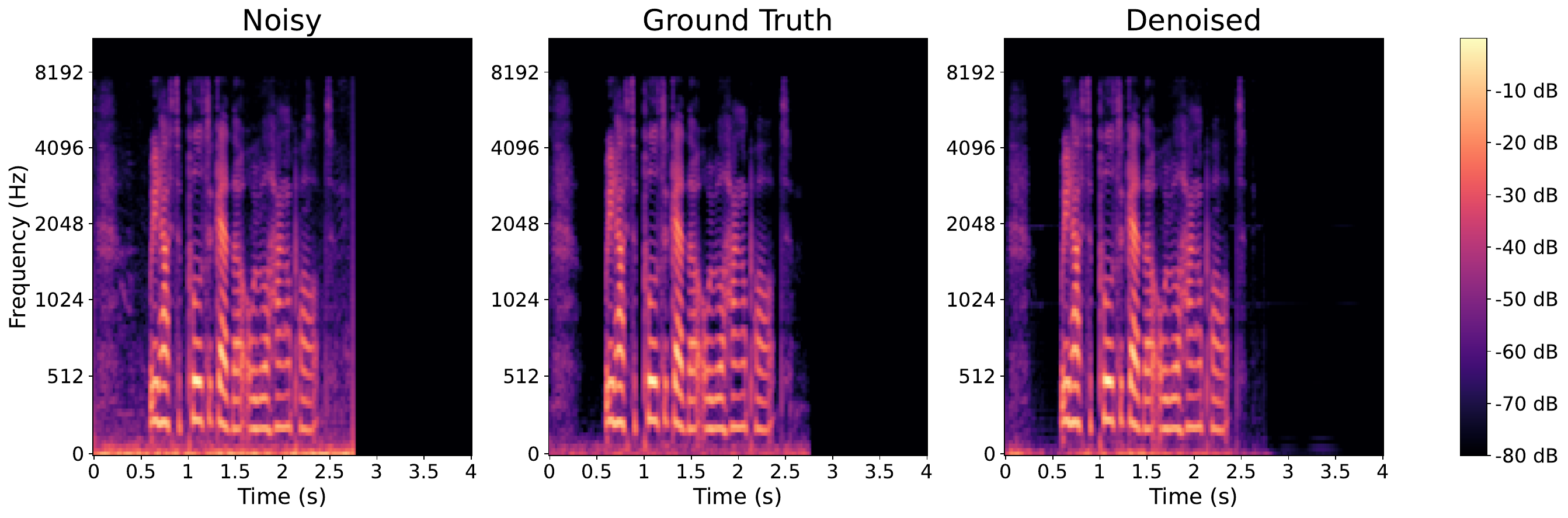}
    \caption{\label{fig:spectral} A comparison of the spectrograms among the noisy signal, the clean ground truth, and the denoised output of a sample in the DNS1 synthetic testset (no reverb). Besides a minor low-frequency artifact in the silent region, the denoised output matches very close to the ground truth signal, despite not using any pre/post-processing in the spectral domain.}
\end{figure}

To further inspect the quality of the denoised samples produced by the network, we listened to the denoised samples from both synthetic data and real recordings, and performed internal A/B tests against other networks. In the supplementary material, we also included the script for generating denoised audios for real recordings from the DNS1 challenge with natural reverberations. In addition, we provide a comparison of the denoised spectrogram and the clean spectrogram in Fig.~\ref{fig:spectral}, to ensure that the denoised samples do not contain any unnatural artifacts as common with raw audio processing systems\footnote{For example, the network may generate low-frequency artificial signals, resulting in a background ``humming" noise. It may also attempt to aggressively attenuate high-frequency speech components, resulting in a robotic tone of speech.}, which may not be captured by the PESQ score \cite{pesqetarian}. In addition, we also tried to test a homogeneous variant of the network working with spectral features (STFT + iSTFT) with a similar number of parameters and MACs, but the PESQ results were significantly worse. 
\ifready
    See Appendix~\ref{app:stft} for more details.
\fi

In addition to audio denoising, we also perform studies on the ability of our network to perform super-resolution and de-quantization on highly compressed data. This involves intentionally performing down-sampling and quantization of the input signals, in that order, and re-training the network to handle the degraded/compressed inputs\footnote{It is possible to perform parameter-efficient fine-tuning of a baseline model instead, such as using low-rank adaptations on the state-space matrices. We leave this to future work.}. To use the same network architecture to interface with down-sampled audios, we perform interleaved repeats of the input signals to restore the original sample rate\footnote{An alternative is to simply remove an encoder block with the same downsampling factor as the downsampled input.}. For quantization, we perform mu-law encoding on the input signals down to the desired bitwidth, then rescale the quantized signal back to the -1 to +1 range. The super-resolution and de-quantization results are reported in Table.~\ref{table:degrade}. Note that the outputs are still evaluated against clean signals at 16000 Hz and full precision.

\begin{table}[htbp]
    \centering
    \caption{\label{table:degrade} The average PESQ scores of the model outputs when the noisy inputs are down-sampled and quantized.}
    \begin{tabular}{lccc}
        \toprule
        \textbf{Input Type} & \textbf{VBDMD} & \textbf{DNS1} \\
        \midrule
        \textbf{8000 Hz \& 8 bit} & 3.19 & 2.88 \\
        \midrule
        \textbf{4000 Hz \& 8 bit} & 3.04 & 2.72 \\
        \midrule
        \textbf{8000 Hz \& 4 bit} & 2.90 & 2.55 \\
        \midrule
        \textbf{4000 Hz \& 4 bit} & 2.72 & 2.39 \\
        \bottomrule
    \end{tabular}
\end{table}

\section{Future Directions}

To make the network even more mobile-friendly, we plan to explore the sparsification and quantization of the network weights and activations. Potentially, we can study a low-rank realization of the state-space matrices, which form the majority of the parameter count. SSMs are also known to be easily adaptable for a spiking implementation on neuromorphic hardware, so it will also be interesting to see whether our model admits an efficient spiking neural network realization, which will further reduce the power required for the solution.

\section{Conclusion}

We introduced a lightweight deep state-space autoencoder, aTENNuate, that can perform raw audio denoising, super-resolution, and de-quantization. Compared to previous works, the key features of this network are: 1) consisting of SSM layers that can be efficiently trained and configured for inference, 2) allowing for real-time inference with low latency, 3) architecturally simple and light in parameters and MACs, 4) capable of processing raw audio waveforms directly without requiring pre/post-processing, and 5) highly competitive with other speech enhancement solutions.

\section{Acknolwedgement}

We thank Temi Mohandespour, Keith Johnson, and Nikunj Kotecha for contributing to the early stages of the project. We also thank M. Anthony Lewis, Douglas McLelland, Kristofor Carlson, and Chris Jones for providing useful feedback on the manuscript.


\ifready
    \appendices

\section{Diagonalization of the $\A$ matrix}
\label{appendix_label}

If we allow $\B$ and $C$ to be complex matrices, then we can assume $\A$ to be diagonal without any loss of generality. To see why, we simply let $\A = P^{-1}\Lambda_AP$ (where $\Lambda_A$ is the diagonalized $\A$ matrix, and $P$ is the similarity matrix) and observe the following:
\begin{equation}
\begin{split}
\forall t, \qquad
k[t] &= C(P^{-1}\Lambda_AP)^{t-1}\B \\
&= C P^{-1} \underbrace{\Lambda_A (P P^{-1}) \, ... \, \Lambda_A (P P^{-1})}_{\text{repeat $t-1$ times}} P \B \\
&= (C P^{-1}) \Lambda_A^{t-1} (P \B) \\
&= C' \Lambda_A^{t-1} \B',
\end{split}
\end{equation}
where $\B'$ and $C'$ are complex matrices that have ``absorbed" the similarity matrix $P$, but WLOG we can just redefine them to be $\B$ and $C$. Since $\A$ is a real matrix, the complex eigenvalues in $\Lambda_A$ must come in conjugate pairs. And WLOG we can again redefine $\Lambda_A$ as $\A$.

\section{Experiment Details}
\label{app:exp_details}

Our baseline model is trained using PyTorch with:
\begin{itemize}
    \item 500 epochs
    \item AdamW optimizer with the PyTorch default configs
    \item A cosine decay scheduler with a linear warmup period equal to 0.01 of the total training steps, updating after every optimizer step
    \item gradient clip value of 1
    \item layer normalization (over the feature dimension) with elementwise affine parameters
    \item SiLU activation
    \item no dropout
\end{itemize}

The high-level training pipeline for the raw audio denoising model is to simply generate synthetic noisy audios by randomly mixing clean and noise audio sources. The noisy sample is then used as input to the model, and the clean sample is used as the target output.

For the clean training samples, we use the processed VCTK and LibriVox datasets that can be downloaded from the Microsoft DNS4 challenge. We also use the noise training samples from the DNS4 challenge as well, which contains the Audioset, Freesound, and DEMAND datasets. For all audio samples, we use the \texttt{librosa} library to resample them to 16 kHz and load them as numpy arrays. 

For the LibriVox audio samples which form long continuous segments of human subjects reading from a book, we simply concatenate all the numpy arrays, and pad at the very end such that the array can be reshaped into $(\text{segments},\, 2^{17})$. For all the other audio samples consisting of short disjoint segments, we perform intermediate paddings when necessary, to ensure a single recording does not span two rows in the final array. For audio samples longer than length $2^{17}$, we simply discard them. The input length to our network during training is then also $2^{17}$.

For every epoch, we use the entirety of the VCTK dataset and 10 percent of a randomly sampled subset of the LibriVox dataset. For each clean segment, we pair it with a randomly sampled noise segment (with replacement). The clean and noise samples are added together with an SNR sampled from -5 dB to 15 dB, and the synthesized noisy sample is then rescaled to a random level from -35 dB to -15 dB. Furthermore, we perform random temporal and frequency masking (part of the SpecAugment transform) on only the input noisy samples.

For the loss function, we combine SmoothL1Loss with spectral loss on the ERB scale, without any equalization procedure on the raw waveforms or spectrograms. The $\beta$ parameter of the SmoothL1Loss is set at 0.5, and the spectral loss is weighted by a factor that grows from 0 to 1 linearly during training. We use a learning rate of 0.005 and a weight decay of 0.02.

\section{Network Details}
\label{app:network}

\subsection{Initialization of SSM Parameters}

As mentioned in Section \ref{ssm}, we make the SSM parameters $\{A, B, C, \Delta\}$ trainable. Recall that we take $A \in \mathbb{C}^h$ to be a complex vector (or a complex diagonal matrix). For stability, we treat the real and complex parts of $A$ separately. The real part $\Re(A)$ is parameterized as $-\text{softplus}(a_r)$, where $a_r$ is initialized with the value $-0.4328$, giving $\Re(A) = -1/2$ initially. Note that due to the positivity of softplus, $\Re(A)$ will always remain negative during training, which ensures the stability of the SSM layer. The imaginary part $\Im(a)$ is parameterized directly and initialized with $\pi(i-1)$ where $i$ is the state index. The matrix $B \in \mathbb{R}^{h \times n}$ is initialized with all ones, and the matrix $C \in \mathbb{R}^{m \times h}$ is initialized with Kaiming normal random variables (assuming a fan in of $h$). Finally, we initialize $\Delta$ with $0.001 \times 100^{\lfloor i / 16 \rfloor}$, giving a series of geometrically spaced values from 0.001 to 0.1 in blocks of 16. These initializations and parameterizations are not absolutely required, and we suspect that any reasonable approach respecting the stability of the SSM layer will suffice.

\subsection{Optimal Contraction Order}

If we let $K(t) = \A^t$ be the ``basis kernels" of the SSM layer, and its Fourier transform be $\hat{K}(f)$, then in einsum form, the SSM layer operations during training can be expressed as 
\begin{equation}
\label{einsum}
\hat{y}_{bjf} = \hat{x}_{bif} \B_{ni} \hat{K}_{nf} C_{jn}
\end{equation}
by virtue of the convolution theorem, where $\{b, i, j, n, f\}$ indexes the batch size, input channels, output channels, internal states, and Fourier modes respectively. With abuse of notation, we similarly let $\{B, I, J, N, F\}$ be the sizes of the five aforementioned dimensions. Note that the number of Fourier modes $F$ is the same as the length of the signal $L$ (or roughly half of $L$ for real FFT).

There are two main ways to compute $\hat{y}$. First, we can perform the operations of Eq.~\ref{einsum} from left to right normally, corresponding to projecting the input, performing the FFT convolution, then projecting the output. Alternatively, we can compute the full kernel, then perform the full FFT convolution with the input as $\hat{x}_{bif}(\B_{ni} \hat{K}_{nf} C_{jn})$. If we only focus on the computational requirements of the forward pass, then the first contraction order will result in $BNIF + BNF + BJNF \approx BNF(I+J)$ units of computation. The second contraction order will result in $JNI + JNIF + BJIF \approx JIF(B+N)$ units of computation. Therefore, we see that the optimal contraction order is intimately linked with the dimensions of the tensor operands. More formally, the first order of contraction is more optimal only when $BNF(I+J) < JIF(B+N)$ or $\frac{1}{B} + \frac{1}{N} > \frac{1}{I} + \frac{1}{J}$.

\subsection{Network Latency}

As mentioned in Section~\ref{sec:network} and Fig.~\ref{fig:network} of the main text, our model is an autoencoder network with long-range skip connections between the encoder and decoder blocks. Each SSM layer retains the length and channels of the input features, and a resampling layer performs both the temporal resampling and channel projection operations. For example, the first SSM block maps the input features as $(1, L) \to (1, L)$, followed by a resampling layer that maps the features as $(1, L) \to (16, L/4)$, through the process described in Section~\ref{sec:network}. The resampling factor and output channel of each block are reported in Table~\ref{tab:network}. For all blocks, the hidden state of the SSM layer is fixed to $h=256$.

If the network does not have any (non-causal) convolutional layers (or PreConvs), then the theoretical latency can be simply determined from the product of the resampling factors, since the ``stride sizes" is the same as the ``kernel sizes" for these layers. In the case of the network given in Table~\ref{tab:network}, the product of the resampling factors (or the total resampling factor) is 256, and the sample rate of the input signal is 16000Hz, so the latency of the network is at least $256 / 16000\text{Hz} = 16$ milliseconds\footnote{Technically, it should be $(256 - 1) / 16000\text{Hz} = 15.9\text{ms}$, but this makes little difference.}. In the presence of PreConvs, or centered convolutional layers of kernel sizes of 3, there are additional latencies introduced due to the non-causal nature of the convolutions. More specifically, the additional latency is the look-forward timestep of the kernel, or simply the step size of the input features.



\section{Importance of raw audio features}
\label{app:stft}

The aTENNuate network receives raw audios as inputs and produces raw audios as outputs directly, which gives rise to the natural hourglass macro-architecture where downsampling and upsampling operations are progressively performed. If we were to opt for a more traditional approach, the network would receive audio features that underwent STFT and produce audio features that will undergo iSTFT. 

A typical window size and hop length for these spectral transform operations are 512 samples (32 ms) and 256 samples (16 ms), respectively. This would effectively yield 257 complex features as input to the network (up to the Nyquist frequency). If we omit the DC component (which can be ``absorbed'' into the biases of the network), then we are effectively left with 256 complex channels. Naturally, we can then build a homogeneous variant of the aTENNuate network by simply stacking the bottleneck blocks, which also expect channels of 256. Here, we choose to repeat the block 12 times, with nested skip connections as prescribed in the original network.

In Table ~\ref{table:stft}, we see that despite requiring more parameters and MACs, the homogeneous network working with spectral features did not outperform the aTENNuate network working with raw audios.

\begin{table}[htbp]
    \centering
    \caption{\label{table:stft} Comparing the aTENNuate raw-audio denoising network against a 12-layer homogeneous S5 network working with spectral features. Both networks contain no PreConv layers, and has a network latency of 16 ms. Note the computational costs of STFT and iSTFT operations are not accounted for.}
    \begin{tabular}{lccccc}
        \toprule
        \textbf{Model} & \textbf{PESQ} & \textbf{Parameters} & \textbf{MACs / sec} \\
        \midrule
        {\bf aTENNuate} & 3.06 & 0.84M & 0.33G \\
        aTENNuate (STFT + iSTFT) & 2.72 & 1.58M & 1.59G \\
        \bottomrule
    \end{tabular}
\end{table}
\fi

\end{document}